%% file: main.tex
\documentclass[sigconf, screen]{acmart}

\input{math_commands.tex}

\usepackage{hyperref}
\usepackage{url}
\usepackage{times}
\usepackage{latexsym}
\usepackage{verbatim}
\usepackage[T1]{fontenc}
\usepackage[utf8]{inputenc}
\usepackage{microtype}
\usepackage{algorithm} 
\usepackage{algpseudocode}
\usepackage{balance}
\usepackage{breqn}
\usepackage{lipsum,graphicx}
\usepackage{booktabs}
\usepackage[export]{adjustbox}
\usepackage{layouts}

\usepackage{caption}
\usepackage{subcaption}
\usepackage{multirow}
\usepackage{makecell}
\usepackage{pifont}
\usepackage{geometry} 
\usepackage{tikz} 
\usepackage{lipsum}

\newcommand{\mybox}[4]{
\vspace{4pt}
\noindent{}\fbox{\parbox{0.963\columnwidth}{\textbf{#1:} 
#4
}
}
\vspace{4pt}

}

\newcommand\jdime{\textsc{JDime}}
\newcommand\jsfstmerge{\textsc{jsFSTMerge}}
\newcommand\fstmerge{\textsc{FSTMerge}}
\newcommand\thistool{MergeBERT}
\newcommand{\ic}[1]{\begin{small}\texttt{#1}\end{small}}

\newcommand{\rqOne}{How effective is \thistool{} in producing merge conflict resolutions?}
\newcommand{\rqTwo}{How well does \thistool{} perform across different languages?}
\newcommand{\rqThree}{How do different choices of context encoding impact performance of \thistool{}?}
\newcommand{\rqFour}{How do users perceive \thistool{} resolutions?}

\title{Program Merge Conflict Resolution via Neural Transformers}

\author{Alexey Svyatkovskiy}
\affiliation{%
  \institution{Microsoft}
  \city{Redmond}
  \state{WA}
  \country{USA}
}

\author{Sarah Fakhoury}
\affiliation{%
  \institution{Washington State University}
  \city{Pullman}
  \state{WA}
  \country{USA}
}

\author{Negar Ghorbani}
\affiliation{%
  \institution{UC Irvine}
  \city{Irvine}
  \state{CA}
  \country{USA}
}

\author{Todd Mytkowicz}
\affiliation{%
  \institution{Microsoft Research}
  \city{Redmond}
  \state{WA}
  \country{USA}
}

\author{Elizabeth Dinella}
\affiliation{%
  \institution{University of Pennsylvania}
  \city{Philadelphia}
  \state{PA}
  \country{USA}
}

\author{Christian Bird}
\affiliation{%
  \institution{Microsoft Research}
  \city{Redmond}
  \state{WA}
  \country{USA}
}

\author{Jinu Jang}
\affiliation{%
  \institution{Microsoft}
  \city{Redmond}
  \state{WA}
  \country{USA}
}

\author{Neel Sundaresan}
\affiliation{%
  \institution{Microsoft}
  \city{Redmond}
  \state{WA}
  \country{USA}
}

\author{Shuvendu K. Lahiri}
\affiliation{%
  \institution{Microsoft Research}
  \city{Redmond}
  \state{WA}
  \country{USA}
}

\renewcommand{\shortauthors}{Svyatkovskiy, Fakhoury, Ghorbani, Mytkowicz, Dinella, Bird, Jang, Sundaresan, Lahiri}

\begin{document}

\begin{abstract}
Collaborative software development is an integral part of the modern software development life cycle, essential to the success of large-scale software projects. When multiple developers make concurrent changes around the same lines of code, a merge conflict may occur. Such conflicts stall pull requests and continuous integration pipelines for hours to several days, seriously hurting developer productivity. To address this problem, we introduce \thistool{}, a novel neural program merge framework based on token-level three-way differencing and a transformer encoder model. By exploiting the restricted nature of merge conflict resolutions, we reformulate the task of generating the resolution sequence as a classification task over a set of primitive merge patterns extracted from real-world merge commit data. Our model achieves 63--68\% accuracy for merge resolution synthesis, yielding nearly a 3$\times$ performance improvement over existing semi-structured, and 2$\times$ improvement over neural program merge tools. Finally, we demonstrate that \thistool{} is sufficiently flexible to work with source code files in Java, JavaScript, TypeScript, and C\# programming languages.
To measure the practical use of \thistool{}, we conduct a user study to evaluate \thistool{} suggestions with 25 developers from large OSS projects on 122 real-world conflicts they encountered. Results suggest that in practice, \thistool{} resolutions would be accepted at a higher rate than estimated by automatic metrics for precision and accuracy. Additionally, we use participant feedback to identify future avenues for improvement of \thistool{}.

\end{abstract}



\setcopyright{acmcopyright}
\acmPrice{15.00}
\acmDOI{10.1145/3540250.3549163} 
\acmYear{2022}
\copyrightyear{2022}
\acmSubmissionID{fse22main-p1294-p}
\acmISBN{978-1-4503-9413-0/22/11}
\acmConference[ESEC/FSE '22]{Proceedings of the 30th ACM Joint European Software Engineering Conference and Symposium on the Foundations of Software Engineering}{November 14--18, 2022}{Singapore, Singapore}
\acmBooktitle{Proceedings of the 30th ACM Joint European Software Engineering Conference and Symposium on the Foundations of Software Engineering (ESEC/FSE '22), November 14--18, 2022, Singapore, Singapore}

\maketitle

\input{introduction.tex}

\input{motivation}

\input{datadriven_merge}

\input{ml.tex}

\input{research-questions.tex}

\input{dataset.tex}

\input{evaluation_combined}

\input{survey}

\input{related_work.tex}

\input{threats}

\input{conclusion.tex}

\bibliography{refs}
\bibliographystyle{ACM-Reference-Format}

\end{document}

%% file: math_commands.tex
\usepackage{amsmath,amsfonts,bm}

\def\eqref#1{equation~\ref{#1}}

\def\1{\bm{1}}

\DeclareMathAlphabet{\mathsfit}{\encodingdefault}{\sfdefault}{m}{sl}
\SetMathAlphabet{\mathsfit}{bold}{\encodingdefault}{\sfdefault}{bx}{n}



%% file: introduction.tex
\section{Introduction}

Collaborative software development relies on version control systems such as \texttt{git} to manage and track changes across a codebase. In most projects, developers work primarily in a branch of a software repository, periodically synchronizing their code changes with the \texttt{main} branch via merges and pull requests~\citep{gousios2016work}. When multiple developers make concurrent changes to the same line of code, a merge conflict may occur.
Merge commits occur frequently, almost 12\% of all commits are related to a merge~\cite{ghiotto2018nature}, and up to 46\% of those commits result in conflicts. Resolving merge conflicts is a time-consuming, complicated, and error-prone activity~\cite{bird2012assessing}. 
To resolve a conflict, developers must stop their workflow, understand conflicting changes, and identify a correct resolution. The ideal way to resolve a conflict is not always clear, and may require referring to project specification documentation or communicating with their peers about changes~\cite{brun2011proactive,de2019recommending,nelson2019life,guimaraes2012improving,bird2012assessing}. 

Modern version control systems such as \texttt{git} utilize the \texttt{diff3} algorithm for performing unstructured line-based three-way merge of input files~\citep{smith-98}. Thus, it is the \emph{de facto} tool for merging and identifying merge conflicts in software development.
This algorithm aligns the two-way diffs of two versions of the code, $\mathcal{A}$ and $\mathcal{B}$, with the common base, $\mathcal{O}$, into a sequence of diff ``slots''.  
At each slot, a change from either $\mathcal{A}$ or $\mathcal{B}$ is selected. 
In cases where both $\mathcal{A}$ \textbf{and} $\mathcal{B}$ contain changes (relative to $\mathcal{O}$) in the same slot (e.g., on the same line), there is a merge conflict.  Standard merge algorithms cannot automatically determine the correct way to merge these conflicting changes. In these cases, developers must manually intervene in order to correctly resolve the conflicting code and complete the merge. 

Over the past decade, several approaches have been proposed to improve the detection and automatic resolution of merge conflicts~\cite{mens2002state,apel2010semistructured,lessenich2017renaming,sousa2018verified,cavalcanti2017evaluating,zhu2018conflict,kasi2013cassandra,brun2011proactive}. Some approaches use the abstract syntax trees (ASTs) or other representations of the source code to improve conflict resolution~\cite{westfechtel1991structure,apel2010semistructured, tavares2019semistructured}; others use a data-driven approach which uses deep learning to predict the correct merge~\cite{Dinella2021}. Researchers have also developed tools to help developers visualize and navigate the merge conflict resolution process~\cite{shen2019intellimerge,semanticmerge,beyondcompare}, and identified key needs of the developer community for effective tool support~\cite{nelson2019life}. The sheer body of research dedicated to this problem represents a significant amount of time and effort. Despite these advancements, none of these approaches have been widely adopted into practice, and the git textual-based detection algorithm remains one of the most commonly used merging approaches~\cite{nelson2019life}.  

\input{example_fig}

In an effort to address this, we introduce \thistool{}: a neural program merge framework based on token-level three-way differencing and a multi-input variant of the bidirectional transformer encoder (BERT) model~\cite{bert}. We formulate the task of generating a merge conflict resolution sequence as a classification task over a set of primitive merge patterns extracted from real-world merge commit data. \thistool{} encodes all inputs that a standard \texttt{diff3} algorithm takes (two two-way diffs of input programs) as well as the edit sequence information, then aggregates them for learning.
We train and then evaluate \thistool{} on 220,000 and 54,000 (respectively) real world historical merge conflicts and their associated manual resolutions from 100,000 GitHub repositories in JavaScript, TypeScript, Java and C\#, and find that it performs quite well, with precision and accuracy always over 60\% (over 70\% if the top three suggestions are considered).  
Further, we compare \thistool{} to existing state of the art structured and semi-structured merge approaches (which are necessarily language-specific) and show that \thistool{} is able to provide resolution suggestions for more merge conflicts and the suggestions are correct (i.e., match the historical user manual resolution) more often.

To better evaluate the resolutions generated by \thistool{} from users' perspective in practice, we also conduct a user study with 25 developers from large OSS projects. We ask participants to evaluate if \thistool{} resolution suggestions are acceptable on a set of 122 of their {\it own} real-world conflicts. Results show that \thistool{} merge resolutions would be accepted in practice despite not always being syntactically identical to the historical user resolutions, and we identify potential ways to improve \thistool{} and the merge conflict oracles used to evaluate neural program merge approaches.

We make the following contributions in this paper:
\begin{enumerate}
    \item We introduce \thistool{}, a novel transformer-based program merge framework that leverages token-level three-way differencing and formulates the task of generating the resolution sequence as a classification task.
    \item We evaluate \thistool{} against structured and semi-structured program merge tools like \jsfstmerge{} and \jdime{}, as well as neural program merge models~\cite{Dinella2021}. We demonstrate that \thistool{} outperforms the state-of-the-art, achieving 2--3$\times$ higher accuracy on merge resolution.
    \item We present an empirical evaluation of the perceptions of \thistool{} resolutions with 25 developers from large OSS projects, contributing the first user study in which developers use and evaluate an automatic merge conflict resolution tool on their own real-world conflicts.
\end{enumerate}

We make available an online data package~\citep{ICSE22Replication} containing the test dataset of conflicts and user resolutions, as well as, the questions and responses gathered from our user study.  We also provide an online Appendix with supplementary details and figures~\cite{FSE22Appendix} (also uploaded with this submission).

%% file: example_fig.tex
\newcommand{\rulesep}{\unskip\ \vrule\ }


\begin{figure*}[t]
    \centering
    \begin{subfigure}[t]{0.3\textwidth}
        \includegraphics[width=\textwidth]{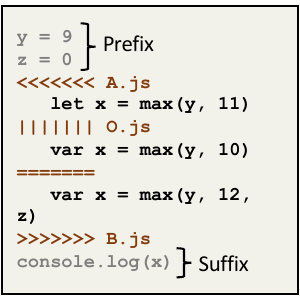}
        \caption{Line-level conflict}
        \label{fig:line-level-conflict-b}
    \end{subfigure}
    \begin{subfigure}[t]{0.3\textwidth}
        \includegraphics[width=\textwidth]{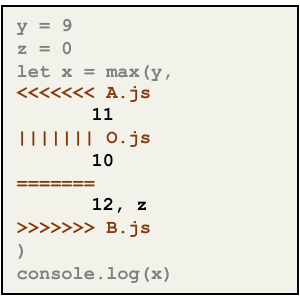}
        \caption{Token-level conflict}
        \label{fig:token-level-conflict-b}
    \end{subfigure}
    \begin{subfigure}[t]{0.3\textwidth}
        \includegraphics[width=\textwidth]{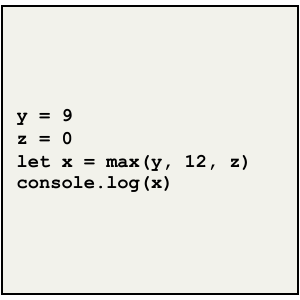}
        \caption{Resolved merge}
        \label{fig:suggested-merge-res-a}
    \end{subfigure}
    \caption{Example merge conflict represented through standard \texttt{diff3} (left) and token-level \texttt{diff3} (center), and the user resolution (right). The merge conflict resolution takes the token-level edit $b$.}
    \label{fig:word1}
\end{figure*}

%% file: motivation.tex
\section{Motivating Example}

We use a number of terms, concepts, and ideas throughout this paper.  To provide an intuition around how our approach works and concretely define terms and concepts, we begin with a motivating example of a small, but realistic merge conflict.

Fig.~\ref{fig:word1} provides an example merge conflict in JavaScript which shows the result of merging two concurrent changes to the same JavaScript file. Fig.~\ref{fig:word1}(a) on the left shows the standard \texttt{diff3} markers ``\texttt{<{}<{}<{}<{}<{}<{}< A.js}'', ``\texttt{||||||| O.js}'', ``\texttt{=======}'' and \\ ``\texttt{>{}>{}>{}>{}>{}>{}> B.js}'', which demarcate the conflicting regions introduced by programs $\mathcal{A}$, base $\mathcal{O}$, and $\mathcal{B}$ respectively. Here, $\mathcal{O}$ represents the lowest common ancestor of programs $\mathcal{A}$ and $\mathcal{B}$ in the version control history. We denote the program text of \texttt{diff3} conflicting regions as $A$, $B$, $O$. The program text outside the conflicting regions – prefix and suffix -- is common to all three programs versions.
Normally conflicts files have the same name in different branches, but to avoid confusion, we name the original file in our example \texttt{O.js}, and the two concurrently edited versions of this file \texttt{A.js} and \texttt{B.js}.
\texttt{A.js} changes ``\texttt{var x''} to ``\texttt{let x}'' and the \texttt{10} to \texttt{11}, while \texttt{B.js} changes the \texttt{10} to \texttt{11} and also adds an argument \texttt{z}.

\thistool{} attempts to automatically resolve merge conflicts in two phases.  
First, \thistool{} represents each line-level merge conflict instance at the token level which localizes conflicting regions.  
Intuitively, \thistool{} converts the three line-structured source texts into three sequences of tokens (including space and line delimiters), applies the standard \texttt{diff3} algorithm to these token sequences, and then reconstructs the merged document at line level. 
Fig.~\ref{fig:word1}(b) shows the result of applying this token-level merge on Fig.~\ref{fig:word1}(a). 
As a result of token-level merge, the whole ``\texttt{let x = max(y,}'' string is cleanly merged, becoming a part of the program prefix, and ``\texttt{)}'' is prepended to the program suffix.  
Second, \thistool{} invokes an underlying neural model to suggest a resolution via classification for each token-level conflicting region and replaces the conflict region with the suggestion from the model (Fig.~\ref{fig:word1}(c)). 

Observe that the resolution does not consist of any single line from either $A$ or $B$ since both edits modify a common line in the base.
Hence, earlier neural approaches such as \texttt{DeepMerge}~\citep{Dinella2021} that are restricted to picking entire lines from the conflict region would not be able to provide the resolution. 
On the other hand, structured merge techniques (such as \jsfstmerge by~\cite{tavares2019javascript}) cannot resolve the conflict soundly as the conflict appears on a program statement, which leads to side effects (e.g. syntactically incorrect code).

A token-level merge can interleave edits within lines (i.e., tokens in which one edit does not conflict with another are trivially merged). Consider $\mathcal{A}$'s edit of the \texttt{var} to \texttt{let} keyword.  
Such non-conflicting edits suffice to demonstrate the above.
Token-level \texttt{diff3} is a syntactic merge algorithm and therefore cannot guarantee semantic or even syntactic correctness of the merged program. 
However, we observed that in practice, syntactic correctness is preserved the majority of the time (over 97\%).

Likewise, consider the token-level conflict for the \texttt{max} function's arguments: an appropriate model trained on JavaScript should easily deduce that taking the edit from $\mathcal{B}$ (i.e., "11, z")  captures the behavior of $\mathcal{A}$'s edit as well. 
The suggested resolution gives an intuitive demonstration of how \thistool{} turns a complex line-level resolution into a simpler token-level classification problem.

%% file: datadriven_merge.tex
\section{Background: Data-driven Merge}
\label{sec:background}
\citet{Dinella2021} introduced the {\it data-driven program merge} problem as a supervised machine learning problem. 
A program merge consists of a 4-tuple of programs $(\mathcal{A}, \mathcal{B}, \mathcal{O}, \mathcal{M})$, where 
\begin{enumerate} 
\item The base program $\mathcal{O}$ is the lowest common ancestor in the version history for programs $\mathcal{A}$ and $\mathcal{B}$, 
\item \texttt{diff3} produces an unstructured line-level conflict when applied to $(\mathcal{A}, \mathcal{B}, \mathcal{O})$, and 
\item $\mathcal{M}$ is the merged program with the developer resolution, incorporating changes made in  $\mathcal{A}$ and $\mathcal{B}$. 
\end{enumerate}
A merge may have multiple unstructured conflicts, we define a {\it merge tuple} $(A, B, O, M)$, where $A, B, O$ correspond to the conflicting regions in $(\mathcal{A}, \mathcal{B}$, and $\mathcal{O})$, respectively, and $M$ denotes the resolution region.

Given a set of merge tuples $(A_i, B_i, O_i, M_i)$, i = 0...N, the goal of a data-driven merge algorithm is to learn a function, $\texttt{merge}$, that maximizes $\sum_{i=0}^{N}\texttt{merge}(A_i, B_i, O_i) = M_i$.
Throughout the text, we will use notations $(a, b, o, m)$ to refer to the token-level merge tuples. 

\citet{Dinella2021} also provide an algorithm for extracting the exact resolution regions for each merge tuple and define a dataset that corresponds to {\it non-trivial} resolutions; resolutions where the developer does not drop the changes from one side of the merge.  
Further, they provide a sequence-to-sequence encoder-decoder based architecture, where a bi-directional gated recurrent unit (GRU) is used for encoding the merge inputs comprising of $(A, B, O)$ segments of a merge tuple, and a {\it pointer mechanism} is used to restrict the output to only choose from line segments present in the input. 
Their paper suffers from two limitations.
First, given the restriction on copying only lines from inputs, their dataset  did not consider merges where the resolution required token-level interleaving, such as the conflict in Figure~\ref{fig:word1}. 
Second, their dataset consists of merge conflicts in a single language, namely JavaScript. 
Our approach addresses both of these limitations.

%% file: ml.tex
\input{mergebert_fig}

\section{Merge Conflict Resolution as a Classification Task}
\label{formulation}

In this work, we demonstrate how to exploit the restricted nature of merge conflict resolutions -- compared to an arbitrary program repair -- to leverage discriminative models to synthesize the merge resolution sequence.
We have empirically observed that the application of \texttt{diff3} at token granularity enjoys two useful properties over its line-level counterpart: (i) it helps localize the merge conflicts to small program segments, effectively reducing the size of conflicting regions, and (ii) most resolutions of merge conflicts produced by token \texttt{diff3} consist entirely of changes from $a$ or $b$ or $o$ or a sequential composition of $a$ followed by $b$ or vice versa. Here, and throughout the paper we will use lower case notations to refer to attributes of token-level differencing (e.g. $a$, $b$, and $o$ are conflict regions produced by \texttt{diff3} at token granularity).
On the flip side, a token-level merge can introduce many small conflicts. 
To balance the trade-off, we start with the line-level conflicts as produced by the standard \texttt{diff3} and perform a token-level merge of only the segments present in the line-level conflict.
There are several potential outcomes for such a two-level merge at the line-level: 
\begin{itemize}
\item {\it A conflict-free token-level merge}: For example, the edit from $A$ about \texttt{let} is merged since $B$ does not edit that slot as shown in Fig.~\ref{fig:word1}(b).  
\item {\it A single localized token-level merge conflict}: For example, the edit from both $A$ and $B$ for the arguments of \texttt{max} yields a single conflict as shown in Fig.~\ref{fig:word1}(b).
\item {\it Multiple token-level conflicts}: Such a case (not illustrated above) can result in several token-level conflicts. 
\end{itemize}

Token-level diff3 applied to a 4-tuple of programs $(\mathcal{A}, \mathcal{B}, \mathcal{O}, \mathcal{M})$, would usually result in a set of localized merge tuples $\langle a_j, b_j, o_j, m_j\rangle$. 
We empirically observe that 74\% of such resolutions $m_j$ are comprised of ($i$) exactly the tokens in $a_j$ or ($ii$) exactly the tokens in $b_j$.  Another 0.4\% of the resolutions are ($iii$) just the tokens in $o_j$. In addition, 23\% of the resolutions are the result of concatenating ($iv$) $a_j$ and $b_j$ or ($v$) $b_j$ and $a_j$.  Finally, 1.8\% comprise another four variants, obtained by taking $i$, $ii$, $iv$ and $v$ above and removing the tokens that also occur in the base, $o_j$. In total, this provides \textit{nine} primitive merge resolution patterns (see online Appendix~\cite{FSE22Appendix} for more details about the primitive merge patterns). 

We, therefore, treat the problem of constructing merge conflict resolutions $m_j$ as a classification task to predict between these possibilities. It is important to note that although we are predicting simple resolution strategies at the token-level, they may translate to complex resolutions at the line-level. In addition, not all conflicts are resolved by breaking that conflict into tokens and applying these patterns---some resolutions such as those introducing new tokens or reordering tokens are not expressible as a choice at the token-level.  


\section{\thistool{}: Neural Program Merge Framework}
\label{sec:main_model}

\thistool{} is a textual program merge model based on the bidirectional transformer encoder (BERT) model~\cite{bert}.
We refer the reader to CodeBERT~\cite{feng-etal-2020-codebert} for a discussion on applying transformers to code. A transformer, like
a recurrent neural network, maps a sequence of text into a high
dimensional representation, which can later be decoded to solve
downstream tasks. While not originally designed for code, transformers have found many applications in software engineering~\cite{clement2020pymt5,kanade2020learning,svyatkovskiy2020intellicode}

\thistool{} approaches merge conflict resolution as a sequence classification task given conflicting regions extracted with token-level differencing and surrounding code as context. 
The key technical innovation in \thistool{} lies in how it breaks program text into an input representation amenable to learning with a transformer encoder and how it aggregates various input encodings for classification. 

In the standard sequence learning setting there is a single input and single output sequence. In the merge conflict resolution task, there are multiple conflicting input programs and one resolution. To facilitate learning in this setting, we construct \thistool{} as a multi-input encoder neural network, which first encodes token sequences of conflicting programs, then aggregates them into a single hidden summarization state. 

An overview of the \thistool{} model architecture is shown in Fig.~\ref{fig:mergebert}. Given conflicting programs $\mathcal{A}$, $\mathcal{B}$ and $\mathcal{O}$ we first perform tokenization and then repeat the three-way differencing at token granularity. If a conflict still exists in this token-level three-way differencing, we collect the token sequences corresponding to conflicting regions $a$, $b$, and $o$, and compute pair-wise alignments of $a$ and $b$ with respect to the base $o$. Finally, for each pair of aligned token sequences we extract an ``edit sequence'' that represents how to turn the second sequence into the first. The resulting aligned token sequences are fed to the multi-input encoder neural network, while the corresponding edit sequences are consumed as type embeddings. Finally, the encoded token sequences are summarized into a hidden state which serves as input to the classification layer. 

Given a 4-tuple of programs $(\mathcal{A}, \mathcal{B}, \mathcal{O}, \mathcal{M})$ which contains token-level merge tuples $(a_{j}, b_{j}, o_{j}, m_{j})$, j=0...N, \thistool{} models the following conditional probability distribution:
\begin{equation}
    p(m_{j} | a_{j}, b_{j}, o_{j}),
\end{equation}
and consequently, for entire programs:
\begin{equation}
    p(\mathcal{M} | \mathcal{A}, \mathcal{B}, \mathcal{O}) = \prod_{j=1}^{N} p(m_{j} | a_{j}, b_{j}, o_{j})
\end{equation}
Independence of token-level conflicts is a simplifying assumption. However, we observe that in our data set only 5\% of merge conflicts result in more than 1 token-level conflict per line-level conflict.

\subsection{Context Encoding}

\begin{figure}
\begin{center}
    \includegraphics[width=.48\textwidth]{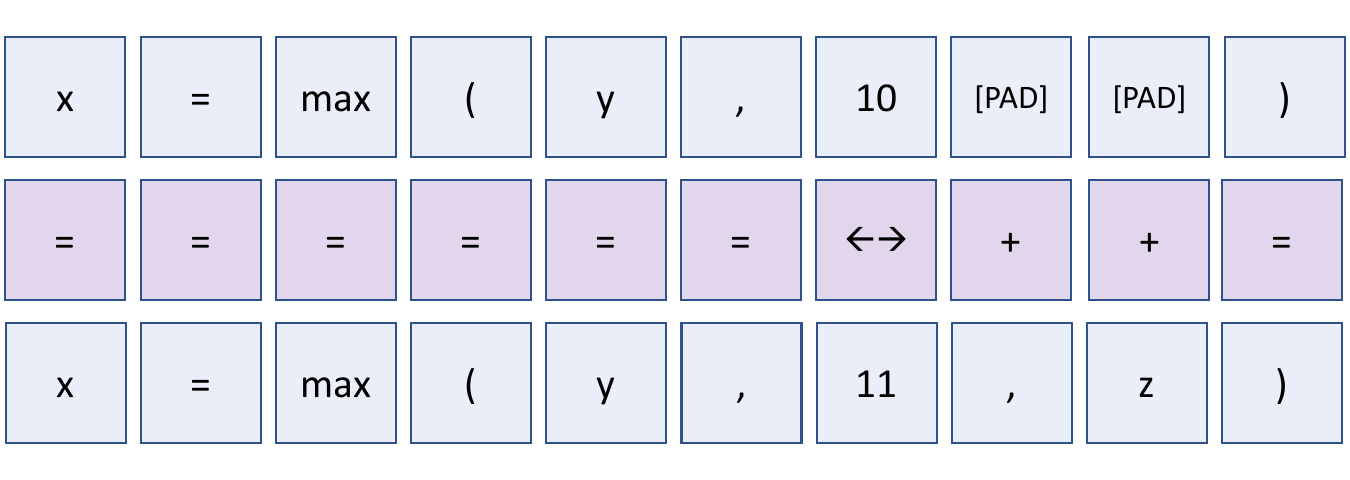}
\caption{An example edit sequence extracted between a pair of token sequences.  Top row is $o|_b$, bottom is $b|_o$, and middle is $\Delta_{bo}$. Padding symbols \texttt{[PAD]} are introduced for alignment. In this case, the target token sequence is obtained by swapping a token and inserting two tokens.}
\label{fig:embedding}
\end{center}
\vspace{-12pt}
\end{figure}

For a merge tuple $(a, b, o, m)$ \thistool{} calculates two pair-wise alignments between the sequences of tokens of conflicting regions $a$ (respectively $b$) with respect to that of the original program $o$: $a|_o$, $o|_a$, $b|_o$, and $o|_b$. For each pair of aligned token sequences we compute an edit sequence. These edit sequences -- $\Delta_{ao}$ and $\Delta_{bo}$ -- are comprised of the following editing actions (kinds of edits): $\textbf{=}$ represents equivalent tokens, $\textbf{+}$ represents insertions, $\textbf{-}$ represents deletions,
$\boldsymbol{\leftrightarrow}$ represents a replacement, and
$\boldsymbol{\emptyset}$ is used as a padding token. Overall, this produces four token sequences and two edit sequences: ($a|_{o}$,
$o|_{a}$, and $\Delta_{ao}$) and ($b|_{o}$, $o|_{b}$, and $\Delta_{bo}$). Fig.~\ref{fig:embedding} provides an example of an edit sequence. Each token sequence covers the corresponding conflicting region and, potentially, surrounding code tokens. We make use of Byte-Pair Encoding (BPE) unsupervised tokenization to avoid a blowup in the vocabulary size given the sparse nature of code identifiers~\cite{10.1145/3377811.3380342}.
To help the model learn to recognize editing steps we introduce an edit type embedding. We combine it with the standard token and position embeddings utilized in BERT model architecture via addition. 

\subsection{Merge Tuple Aggregation}

We utilize transformer encoder model $\mathcal{E}$ to independently encode each of the four token sequences of token-level conflicting regions $a|_{o}$, $o|_{a}$, $b|_{o}$, and $o|_{b}$, passing corresponding edit sequences $\Delta_{ao}$ and $\Delta_{bo}$ as type embeddings. Finally, \thistool{} aggregates the resulting encodings into a single hidden summarization state $h$:
\begin{dmath}
h = \sum_{x \in (a|_{o}, o|_{a}, b|_{o}, o|_{b})} \theta_{x} \cdot \mathcal{E} (x, \Delta_x)
\end{dmath}
where $\mathcal{E} (x, \Delta_x)$ are the encoded tensors for each of the sequences $x \in (a|_{o}, o|_{a}, b|_{o}, o|_{b})$, and $\theta_{x}$ are learnable weights. After aggregation a linear classification layer with \texttt{softmax} is applied:
\begin{equation}
      p(m_{j} | a_{j}, b_{j}, o_{j}) = \mathrm{softmax}(W\cdot h + b)
\end{equation}

The resulting line-level resolution region is obtained by concatenating the prefix, predicted token-level resolution $m_{j}$, and the suffix. Finally, in the case of a one-to-many correspondence between the original line-level and the token-level conflicts (see Appendix for more details and a pseudocode), \thistool{} uses a standard beam-search to decode the most promising predictions. 



\subsection{Implementation Details}
\label{sec:implement}

We utilize a pretrained CodeBERT\footnote{\url{https://huggingface.co/huggingface/CodeBERTa-small-v1}} model with 6 encoder layers, 12 attention heads, and a hidden state size of 768. The vocabulary is constructed using byte-pair encoding \citep{sennrich2015neural} and the vocabulary size is 50000. We transfer the weights of the pretrained transformer encoder into the \thistool{} multi-input neural network, and attach a randomly initialized linear layer with softmax. We then finetune the resulting neural network in a supervised setting for the sequence classification task. Input sequences for finetuning training cover conflicting regions and surrounding code (i.e., fragments of prefix and suffix of a conflicting region) up to a maximum length of 512 BPE tokens. The backbone of our implementation is HuggingFace's~\footnote{\url{https://github.com/huggingface/transformers}} \texttt{RobertaModel} and \\
\texttt{RobertaForSequenceClassification} classes in PyTorch, which are modified to turn the model into a multi-input architecture shown in Fig.~\ref{fig:mergebert}. 
We finetune \thistool{} with Adam stochastic optimizer with weight decay fix using a
learning rate of 5e-5, 512 batch size and 8 backward passes per \texttt{allreduce}. 
The finetuning training was performed on 4 NVIDIA Tesla V100 GPUs with 16GB memory for 6 hours. 

In the inference phase, the model prediction for each line-level conflict consists of one or more token-level predictions. Given the token-level predictions and the contents of the merged file, \thistool{} generates the code corresponding to the resolution region. The contents of the merged file include the conflict in question and its surrounding regions. Afterward, \thistool{} checks the syntax of the generated code with a tree-sitter\footnote{\url{https://tree-sitter.github.io/tree-sitter}} parser and outputs it as the candidate merge conflict resolution only if it is syntactically correct.

%% file: mergebert_fig.tex
\begin{figure*}
\begin{center}
    \includegraphics[width=.85\textwidth]{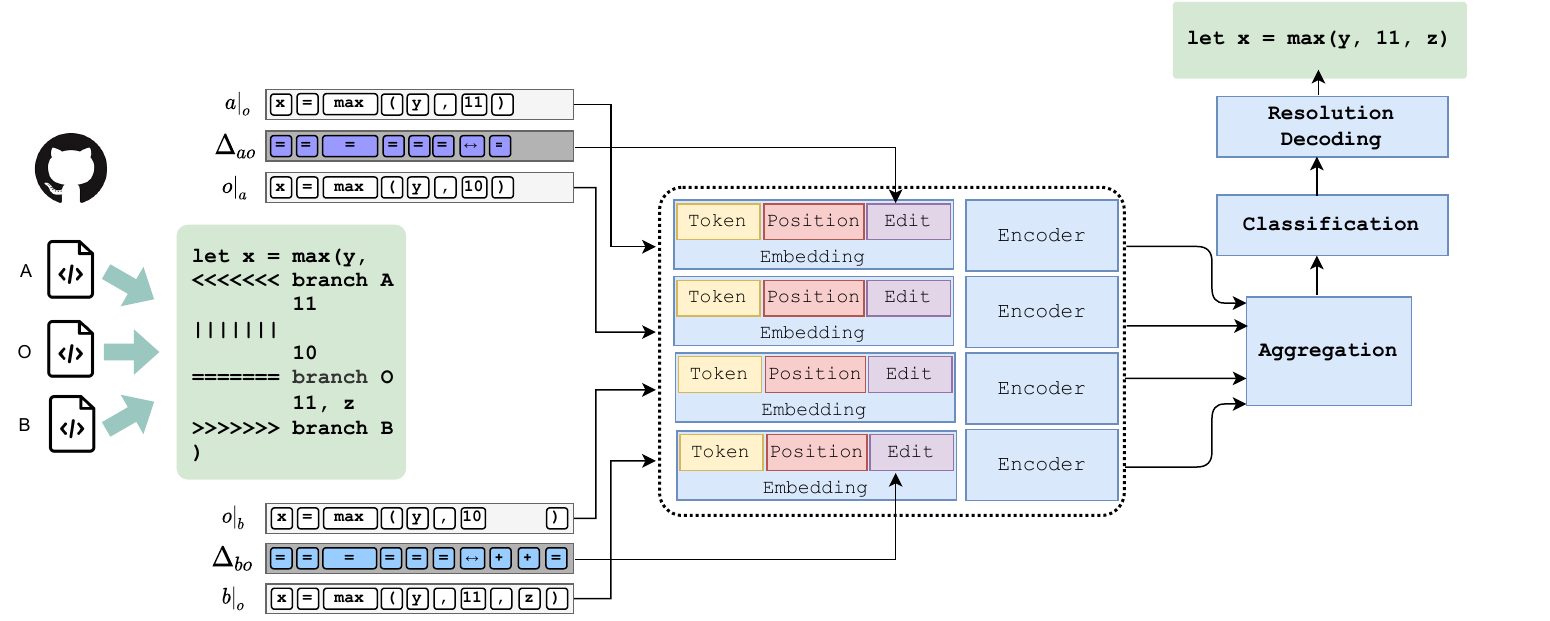}
\caption{An overview of the \thistool{} architecture. From left to right: given conflicting programs $\mathcal{A}$, $\mathcal{B}$ and $\mathcal{O}$ token-level differencing is performed first, next, programs are tokenized and the corresponding sequences are aligned ($a|_o$ and $o|_a$, $b|_o$, and $o|_b$). We extract edit steps for each pair of token sequences ($\Delta_{ao}$ and $\Delta_{bo}$). Four aligned token sequences are fed to the multi-input encoder neural network, while edit sequences are consumed as edit type embeddings. Finally, encoded token sequences are aggregated into a hidden state which serves as input to classification layer.}
\label{fig:mergebert}
\end{center}
\vspace{-8pt}
\end{figure*}

%% file: research-questions.tex
\section{Research Questions}

We pose the following research questions to evaluate the effectiveness of utility of \thistool{}.

\noindent \textbf{RQ\scriptsize{1}: }\textbf{\rqOne}
We evaluate \thistool{}'s performance of producting resolutions in terms of precision and accuracy of matching the actual user resolution extracted from real-world merge resolutions. We also provide a comparison \thistool{} to baseline approaches (at both the line and token level) and state of the art merge resolution approaches.

\noindent \textbf{RQ\scriptsize{2}: }\textbf{\rqTwo}
One of our primary goals is to be able to work on multiple languages with minimal effort.  
The core approach of \thistool{} is fundamentally language agnostic (though a parser and tokenizer is required for each additional language).  
We evaluate performance of \thistool{} across four languages and also compare the results of using four language-specific models (each trained on just one language) to using one multi-lingual model trained on the data from all four languages.

\noindent \textbf{RQ\scriptsize{3}: }\textbf{\rqThree}
We conduct an ablation study of the edit type embedding to understand and evaluate the impact of our novel edit-aware encoding on model performance.

\noindent \textbf{RQ\scriptsize{4}: }\textbf{\rqFour}
We conduct a user study involving a survey of real-world conflicts recently encountered by developers from large OSS projects. To understand how developers would use \thistool{} in practice, we provide them with an interface to explore \thistool{}'s conflict resolution suggestions in relation to their original conflicting code ask them evaluate suggestions and explain why they do or do not correctly resolve the merge conflict.

%% file: dataset.tex
\section{Dataset}
\label{sec:dataset}

The finetuning dataset is mined from over 100,000 open source software repositories in multiple programming languages with merge conflicts. It contains commits from git histories with exactly two parents, which resulted in a merge conflict.  We replay \texttt{git merge} on the two parents to see if it generates any conflicts. Otherwise, we ignore the merge from our dataset. We use the approach introduced by~\citet{Dinella2021} to extract resolution regions---however, we do not restrict ourselves to conflicts with less than 30 lines only.  Lastly, we extract token-level conflicts and conflict resolution classification labels (introduced in Section \ref{formulation}) from line-level conflicts and resolutions. Tab.~\ref{tab:fintuning_dataset} provides a summary of the finetuning dataset.

\begin{table}[htb]
\centering
\caption{Number of merge conflicts in the dataset.}
\begin{tabular}{llllllllllll} \toprule
\textbf{Programming language} & \textbf{Development set}  & \textbf{Test set} \\ \midrule
C\# & 27874 & 6969 \\ 
JavaScript & 66573 & 16644\\ 
TypeScript & 22422 & 5606\\ 
Java & 103065 & 25767 \\ 
\bottomrule
\end{tabular}
\label{tab:fintuning_dataset}
\end{table}
The finetuning dataset is split into development and test sets in the proportion 80/20 at random at the file-level. The development set is further split into training and validation sets in 80/20 proportion at the merge conflict level.

%% file: evaluation_combined.tex
\section{Evaluation}

\subsection{Evaluation Metrics}

We evaluate \thistool{}'s performance of resolution synthesis in terms of precision and accuracy of string match (modulo whitespaces or indentation) to the user resolution extracted from real-world historical merge resolutions. This approach is rather restrictive as a suggested resolution might differ from the actual user resolution by, for instance, only the order of statements, being semantically equivalent otherwise. As such, this evaluation approach gives a lower bound of performance.

We evaluate \thistool{} and compare it to baselines and existing approaches using two metrics, precision at top-k and accuracy at top-k.  
Since \thistool{} is a neural approach, it may provide more than one suggestion, which we rank according to the associated prediction probabilities.
In addition, because we filter out resolution suggestions that are not syntactically valid, it may provide no suggestions in rare cases.  
Accuracy at top-1 indicates the percentage of total conflicts for which \thistool{} produces the correct resolution as its top suggestion. Precision at top-1 indicates how often (as a percentage) the top suggestion is correct when the \thistool{} provides any suggestions at all.  As a concrete example, if a tool produces a resolution suggestion for 50 out of 100 conflicts and 40 of the suggestions matched the actual historical user resolution, then the precision would be 80\% (40/50), but the accuracy would be 40\% (40/100).  Precision at top-k indicates how often the correct resolution is found in the top-k suggestions and Accuracy at top-k is analogous. When ``top-k'' is omitted from the metric name (e.g. just "Precision") then k is 1.



\subsection{Baseline Models}
\label{sec:baselines}

\subsubsection{Language Model Baseline}

Neural language models (LMs) have shown great performance in natural language generation~\citep{gpt2, sellam-etal-2020-bleurt}, and have been successfully applied to the domain of source code~\citep{10.5555/2337223.2337322, gptc, feng-etal-2020-codebert}. We consider the generative pretrained transformer language model for code (GPT-C) and appeal to the naturalness of software~\citep{naturalness} to construct our baseline approach for the merge resolution synthesis task. We establish the following baseline:
given an unstructured line-level conflict produced by \texttt{diff3}, we take the common source code prefix acting as user intent for program merge. We attempt to generate an entire resolution region token-by-token using beam search. As an ablation experiment, we repeat this for the conflicts produced with the token-level differencing algorithm (Fig.~\ref{fig:word1} shows details about prefix and conflicting regions).

\subsubsection{DeepMerge: Neural Model for Interleavings}

Next, we consider \textsc{DeepMerge}~\citep{Dinella2021}: a sequence-to-sequence model based on the bidirectional GRU summarized in section~\ref{sec:background}. It learns to generate a resolution region by choosing from line segments present in the input (line interleavings) with a pointer mechanism. We retrain the \textsc{DeepMerge} model on our TypeScript dataset.

\subsubsection{JDIME}
Looking for a stronger baseline, we consider \textsc{JDime}, a Java-specific merge tool that automatically tunes the merging process by switching between structured and unstructured merge algorithms \citep{apel2012structured}. Structured merge is abstract syntax tree (AST) aware and leverages syntactic information to improve matching precision of conflicting nodes.  We use the publicly available implementation~\citep{jdime}, and run JDime in semi-structured mode. 

\subsubsection{jsFSTMerge}
\citet{tavares2019javascript} implemented \jsfstmerge{} by adapting an off-the-shelf grammar for JavaScript to address shortcomings of \fstmerge{}~\cite{apel2012fstmerge} and modify its algorithm.
\jsfstmerge{} allows for certain types of nodes to maintain their relative order (\emph{e.g.}, statements) while others may be order independent (\emph{e.g.}, function declarations) even when sharing the same parent node.
For cases where \jsfstmerge{} produces a resolution not matching the user resolution, we manually inspect the output for semantic equivalence (e.g., reordered import statements).

\subsection{Results}
\label{sec:eval}

\noindent \textbf{RQ\scriptsize{1}: }\textbf{\rqOne}

To evaluate \thistool{} We first compare it to other neural approaches and to \texttt{diff3}. 
To be comprehensive, we evaluate at both the token level and the line level.  
We then compare \thistool{} to existing state of the art structured and semi-structured merge language-specific merge approaches.

\begin{table}[htb]
\small
\caption{Evaluation results for \thistool{} and various neural baselines calculated for merge conflicts in TypeScript programming language test set. The table shows top-1 precision and accuracy metrics.}
\centering
\begin{tabular}{lllllllllll} \toprule
\textbf{Approach}  & \textbf{Granularity} & {\textbf{Precision}} & {\textbf{Accuracy}} \\ \midrule
LM   & Line  &3.6 & 3.1 \\      
DeepMerge & Line  & 55.0 & 35.1  \\ 
\midrule
\texttt{diff3} & Token & 82.4 & 36.1  \\
\midrule
LM & Token  & 49.7  & 48.1    \\      
DeepMerge & Token  & 64.5 & 42.7 \\ 
\thistool{} & Token  & \textbf{69.1} & \textbf{68.2}  \\  
\bottomrule   
\end{tabular}
\label{tab:baselines_left}
\end{table}

As seen in Tab.~\ref{tab:baselines_left}, language model baselines' performance on merge resolution synthesis is relatively low, suggesting that the naturalness hypothesis is insufficient to capture the developer intent when merging programs. This is perhaps not surprising given the notion of precision that does not tolerate even a single token mismatch. 

\thistool{} is based on two core components: token-level \texttt{diff3} and a multi-input neural transformer model. The token-level differencing algorithm alone gives a high top-1 precision of 82.4\%, with a relatively low accuracy of only 36.1\% (i.e., it doesn't always generate a resolution suggestion, but when it does, it is very often correct). Combined with the neural transformer model, the accuracy is increased to a total of 68.2\%. Note, as a deterministic algorithm token-level \texttt{diff3} can only provide a single suggestion. 

DeepMerge precision of merge resolution synthesis is quite admirable, showing 55.0\% top-1 precision. However, it fails to generate predictions for merge conflicts which are not representable as a line interleaving. This type of merge conflict comprises only roughly one third of the test set, resulting in an accuracy of only 35.1\% which is significantly lower than \thistool{}.


As an experiment, we also evaluate the DeepMerge model in combination with the token-level \texttt{diff3}. This enables DeepMerge to overcome the limitation of providing only resolutions comprised of interleavings of lines from the conflict region by interleaving tokens instead. As seen in Tab.~\ref{tab:baselines_left} (DeepMerge with Token granularity) overall accuracy improves from 35.1\% to 42.7\%. However this still falls short of \thistool{} with precision that is 5\% less (64.5\% vs. 69.1\%) and accuracy that is 25\% less (42.7\% vs 68.2\%). 


\begin{table}[htb]
\small
\caption{Comparison of \thistool{} to \jdime{} and \jsfstmerge{} semi-structured merge tools. The table shows the percentage of conflicts in which the tool produces a resolution, the top-1 precision of produced resolutions, and the overall top-1 accuracy of merge resolution synthesis. \jdime{} evaluation is on a Java data set and \jsfstmerge{} is on a JavaScript data set.}
\vspace{-4pt}
\centering
\resizebox{0.98\columnwidth}{!}{%
\begin{tabular}{lllllll} \toprule
\textbf{Approach} & \textbf{Language} & \textbf{\% conf. w/ res.} & \textbf{Precision} & \textbf{Accuracy} \\ 
\midrule
\jdime{} & Java & 82.1 & 26.3 & 21.6 \\ 
\thistool{} & Java & \textbf{98.9} & \textbf{63.9} & \textbf{63.2} \\ \midrule 
\jsfstmerge & JavaScript & 22.8 & 15.8 & 3.6 \\ 
\thistool{} & JavaScript & \textbf{98.1} & \textbf{66.9} & \textbf{65.6} \\ 
\bottomrule
\end{tabular}%
}
\label{tab:baselines_right}
\vspace{-6pt}
\end{table}

We also compared \thistool{} to state of the art structured and semi-structured merge tools.  Since both \jdime{} and \jsfstmerge{} are language-specific, to compare against \thistool{}, we use our dataset's corresponding language-specific subset of conflicts (leading to slightly different results for \thistool{} on Java and JavaScript).

As can be seen from Tab.~\ref{tab:baselines_right}, \jsfstmerge{} only produces a resolution for 22.8\% of conflicts and when a resolution is produced by \jsfstmerge{}, it is only correct 15.8\% of the time, yielding a total accuracy of 3.6\%. 
This is in line with the conclusions of the creators of \jsfstmerge{} that semi-structured merge approaches may not be as advantageous for dynamic scripting languages~\citep{tavares2019javascript}. Because \jsfstmerge{} may produce reformatted code, we manually examined cases where a resolution was produced but did not match the user resolution (our oracle).  If the produced resolution was semantically equivalent to the user resolution, we classified it as correct.

To compare the accuracy of \textsc{JDime} to that of \thistool{}, we use the Java Test data set introduced previously and complete the following evaluation steps: \textsc{JDime} does not merge all conflicts and generates a resolution for 82.1\% of conflicts. This is in line with related work reporting that as much as 21\% of files cannot be merged~\cite{apel2012structured}. Therefore, first, we identify the set of merge conflict scenarios where \texttt{diff3} reports a conflict and \textsc{JDime} produces a non-conflicted merge. 
When comparing the \textsc{JDime} output to the actual historical user-performed merge conflict resolution, we do not use a simple syntactic match.  As a result of its AST matching approach, code generated by \jdime{} is reformatted, and the original order of statements and other constructs are not always preserved. 
In an effort to accurately and fairly identify semantically equivalent merges, we use GumTree \cite{FalleriMBMM14}, an AST differencing tool, to identify and ignore semantically equivalent differences between \textsc{JDime} output and the user resolution, such as reordered method declarations. When \textsc{JDime} produces a resolution, it generates a semantically equivalent match 26.3\% of the time, with an accuracy of 21.6\%. 

\noindent \textbf{RQ\scriptsize{2}: }\textbf{\rqTwo}
One goal of our approach is to be able to handle multiple languages with minimal effort.  For \thistool{} to be able to provide merge resolution suggestions for conflicts in a particular language, it needs three things.  First, a tokenizer in that language, which allows us to split the source text into tokens for processing.  Second, a parser in that language, which allows us to filter out syntactically incorrect merge resolution suggestions. Third, a data set of merge conflicts and their user-resolutions to train \thistool{}.  Fortunately, tokenizers and parsers for nearly any language are readily available (e.g., we use GitHub's tree-sitter for this) and repositories that use a particular language can be easily identified (e.g. on GitHub) and mined for conflicts and resolutions.

We incorporated tokenizers and parsers into \thistool{} for JavaScript, TypeScript, Java, and C\# and gathered merge conflict data for these languages as described previously.  
Note that both comments and strings in these languages are represented as single tokens and can be quite long.  
Therefore we further split these tokens on whitespace.
Tab.~\ref{tab:mergebert_summary} shows the detailed evaluation results of \thistool{} broken down by language. The top section of results shows performance when \thistool{} is trained on data for that specific language.  
The bottom section shows performance for each language when \thistool{} is trained on a data set comprising data for all languages (we term this the \emph{multilingual} model).
Note that for the language specific models, performance is fairly consistent across all four languages with Top-1 precision ranging from 63.9\% to 69.1\% and Top-1 Accuracy ranging from 63.2\% to 68.2\%. We also find that over 97\% of \thistool{} suggestions are syntactically correct across all programming languages. 

We had no a priori expectations of the performance of the multilingual model, as it is trained on more data, which could lead to improvement, but it is not language specific, which could lead to poorer results.
Overall, the multilingual variant of the model generates results that are just slightly below the monolingual versions.
Thus performance on one language isn't improved by adding more data in other languages.
Thus, from a pragmatic perspective, if one chooses to simplify their use of \thistool{} by training just one model instead of one model per language, then the performance takes only a negligible hit.

\begin{table}
\small
\caption{Detailed evaluation results for (top) monolingual JavaScript, TypeScript, Java, and C\# models, and (bottom) multilingual \thistool{} model trained on all four programming languages. The table shows precision and accuracy of merge resolution synthesis.}
\vspace{-4pt}
\centering
\begin{tabular}{llllllllllll} \toprule
\textbf{Test (Train) Languages} & \multicolumn{2}{c}{\textbf{Precision}} &  \multicolumn{2}{c}{\textbf{Accuracy}}  \\ \cmidrule{2-3} \cmidrule{4-5} 
& Top-1 & Top-3 & Top-1 & Top-3 \\ 
\midrule
JavaScript (JS)  & 66.9 &75.4 & 65.6& 73.9 \\ 
TypeScript (TS)  & 69.1 &76.6 & 68.2& 75.6 \\ 
Java (Java)  & 63.9 &76.1 & 63.2 &75.2 \\ 
C\# (C\#)  & 68.7 &76.4 & 67.3& 74.8 \\ 
\midrule
JavaScript (JS, TS, C\#, Java)  & 66.6& 75.2 & 65.3 &73.8 \\ 
TypeScript (JS, TS, C\#, Java)  & 68.5 &76.8 & 67.6 &75.8 \\ 
Java (JS, TS, C\#, Java)  &  63.6 &76.0 & 62.9& 75.1 \\ 
C\# (JS, TS, C\#, Java)  & 66.3 &76.2 & 65.1 &74.8 \\ 
\bottomrule
\end{tabular}
\label{tab:mergebert_summary}
\vspace{-4pt}
\end{table}

\noindent \textbf{RQ\scriptsize{3}: }\textbf{\rqThree}

We conduct an ablation study on the edit type embedding to understand the impact of edit-awareness of encoding on the model performance. As shown in Tab.~\ref{tab:edit_ablation}, use of the edit type embedding improves  \thistool{} from 63\% to 68\%.
\begin{table}[htb]
\small
\caption{Evaluation results for \thistool{} and the model variant without edit-type embedding for merge conflicts in TypeScript programming language test set. The table shows top-1 precision and accuracy metrics.}
\centering
\begin{tabular}{lllllll} \toprule
\textbf{Approach} & \textbf{Precision} & \textbf{Accuracy}   \\ 
\midrule
w/o edit type embeddings  & 65.2 & 63.1  \\
\thistool{} w/ edit type embeddings & \textbf{69.1} & \textbf{68.2}  \\ 
\bottomrule
\end{tabular}
\label{tab:edit_ablation}
\end{table}


%% file: survey.tex
\section{User Evaluation}
\begin{figure}[h] 
   \includegraphics[width=0.46\textwidth, angle=0]{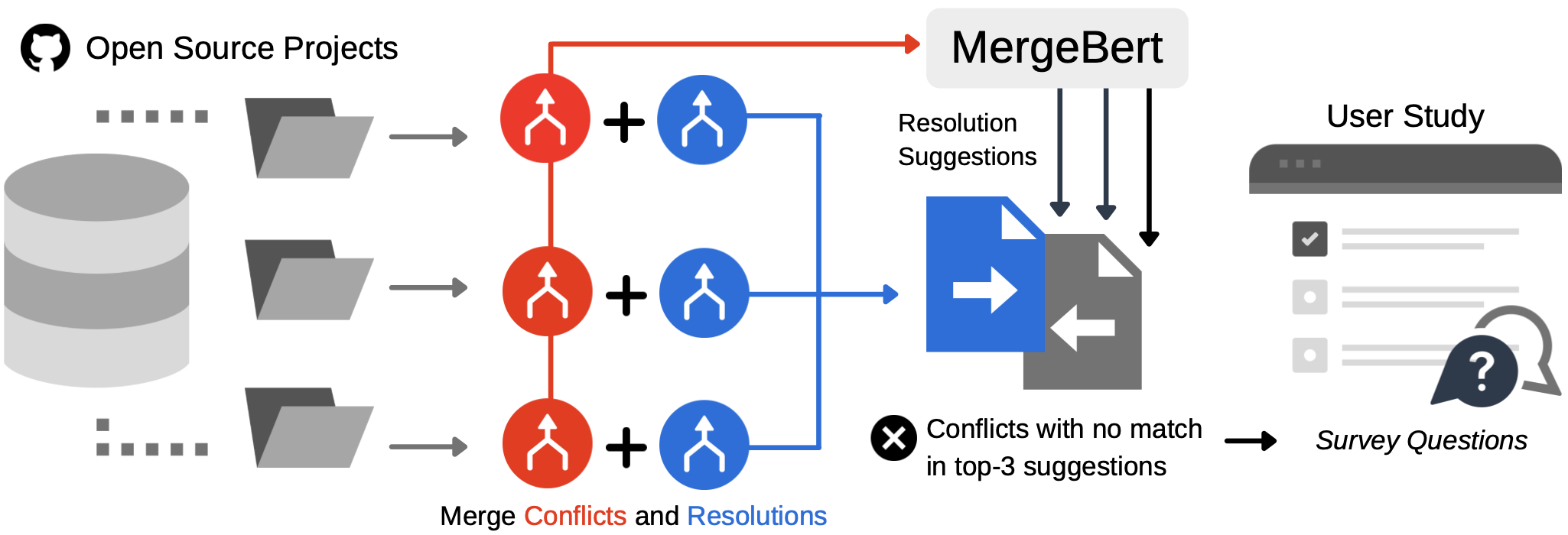}
  \caption{Methodology to identify candidate conflicts for the user study.}
  \label{fig:methodology}
  \vspace{-12pt}
\end{figure}

\begin{table}[t!]
\small
\centering
\caption{Summary of projects in user study, total number of conflicts per project, number of conflicts evaluated in the study, and the survey participants.}
\label{tab:projects}
\begin{tabular}{llcccl} \toprule
Language & Project & Conflicts & Survey & Participants \\
 &  &  &  Conflicts &  \\\midrule
\multirow{3}{*}{Java} & Azure-Cosmosdb  & 341 & 6 & P1 \\
 & Azure-SDK  & 997 & 14 & P2-4 \\
 & ApplicationInsights&  313 & 10 & P5-6 \\ \midrule
\multirow{2}{*}{TS} & MakeCode & 106 & 12 & P7-8 \\
 & VSCode  & 2256 & 48 & \begin{tabular}[c]{@{}l@{}}P9-17\end{tabular} \\ \midrule
\multirow{3}{*}{C\#} & AspNetCore  & 567 & 11 & P18-19 \\
 & EFCore  & 397 & 7 & P20-21 \\
 & Roslyn  & 1894 & 14 & P22-25 \\ \midrule
 Total & 8 projects  & 6871 & 122 & 25 \\ \bottomrule
\end{tabular}
\end{table} 

\subsection{User Study Design}

To better understand how \thistool{} performs in practice, we ask developers about conflicts that \thistool{} is unable to correctly resolve. Since \thistool{}'s resolution suggestions are evaluated against user resolutions using a verbatim string match (modulo whitespace), asking study participants to confirm identical resolutions predicted by \thistool{} is not informative. Therefore, we extract conflicts where \thistool{} suggestions are not a direct match to the user resolution to determine what the limitations of the suggestions are, and how they might be perceived in practice.

To build an oracle of merge conflicts and resolutions we identify 8 open source projects hosted on GitHub. The selected projects are active, with multiple contributors, and contain a large number of conflict scenarios in one of the languages supported by \thistool{}.
Tab.~\ref{tab:projects} contains a list of projects chosen.
For each project, we follow the same steps outlined in Section~\ref{sec:dataset} to extract candidate conflicts and user resolutions to use in the survey.

Fig.~\ref{fig:methodology} explains the methodology used to identify candidate merge conflicts. We identify the set of conflicts \thistool{} is unable to correctly merge (within the top-3 suggestions). From this set of conflicts, we identify candidate conflicts to use as part of the user study. We filter candidate files with the following criteria:
\begin{enumerate}
  \item Conflicts should have been recently resolved i.e., at most within the past 12 months. Participants may not retain the context needed to evaluate suggestions for older conflicts. 
    \item Files must have at most 4 conflicts. Participants evaluate up to 3 suggestions per conflict. More conflicts may be too complex to evaluate within the interview time slot. 
    \item Conflicts should be non-trivial.  Trivial conflicts, such as those that only involve formatting changes or renames, are manually excluded. The determination of if a conflict was non-trivial was manual and subjective, informed by our belief that more substantive conflicts would lead to more insights in the user study.
\end{enumerate}

For each candidate conflict identified, we use the GitHub API to identify authors for each of the conflicting branches and the resolved file. Authors with at least 3 candidate merge conflicts are identified as potential survey participants. Our final pool of candidate participants consists of 52 unique authors. We recruit participants via email, using contact information on GitHub. Out of the 52 contacted developers, 25 agreed to participate in the study. All participants were professional software developers with at least 2-8 years of experience working at large technology companies. We asked participants to evaluate \thistool{} resolution suggestions for their own merge conflicts.  Tab.~\ref{tab:projects} contains the final number of participants and conflicts evaluated in our study. 122 conflicts were evaluated: 32 C\# conflicts, 30 Java, and 60 Typescript. 

\subsubsection{\thistool{} Interface}
We designed an online interface where participants can view their own conflicts and explore \thistool{}'s resolution suggestions. Participants are asked to evaluate their own recently resolved merge conflicts, and the corresponding generated resolution suggestions by \thistool{}. The interface is customized based on the signed-in participant and displays a list of their recently encountered merge conflicts. Participants can click through different resolution suggestions to evaluate if they are acceptable ways to resolve the merge conflict. They can view their original resolution on the same page, and if needed, participants can navigate to the conflicting commit on GitHub using a link if they need additional context. They can also view a diff between the conflict file and any of the selected options (resolution suggestion or user resolution). Participants use this interface to select one or more of the suggested resolutions, indicate if the suggested resolution is acceptable, and explain the reasons why or why not.  Our online data package~\cite{ICSE22Replication} and appendix~\cite{FSE22Appendix} contain the questions, images of the interface, and participant responses.  


\subsubsection{Protocol}
The user study was conducted as 30 minute interviews remotely over Microsoft Teams using the interface we built. First, participants watched a video explaining \thistool{} and how to navigate conflicts and evaluate resolution suggestions using the interface. Then, the participants evaluated a set of conflicts and submitted their responses. One of the authors was on the teams call to help participants navigate the interface and ask any clarifying questions based on their evaluation of the \thistool{} resolution suggestions.
Questions were iteratively developed based on two pilot interviews. Each interview was recorded for transcription and analysis. 


\subsection{User Study Results}

\noindent \textbf{RQ\scriptsize{4}: }\textbf{\rqFour}

Using the interface participants evaluate the conflict resolution suggestions generated by \thistool{} and indicate if any of the suggestions were acceptable, and explain why or why not. There were no noticeable differences in the participants' responses across different languages or projects so we do not break down our results by those dimensions.
Participant's evaluations of the merge suggestions generally fall into three categories: 1) the merge suggestion is correct and would be used to resolve the conflict 2) the merge is incorrect but the correct resolution would require an understanding of external context and 3) the merge is incorrect and no external context is needed.  

\subsubsection{Acceptable Merge Suggestions}

Surprisingly, of the 122 conflicts included in the study, participants indicated that at least one of the 3 suggestions generated by \thistool{} was correct for 54\% (66/122) of the examples. By design, the suggestions presented in the study are not syntactically equivalent to the participant's original resolution, however, they still indicated that the suggestion was a correct merge. Using participant responses, we identify a few reasons why merge suggestions may be acceptable to a developer, even if it is not syntactically equivalent to their original resolution:

\vspace{6pt}
\noindent{}\textbf{Semantically Equivalent Resolution} (54 of 122 conflicts) \\
    Semantically equivalent resolutions include scenarios where the statements are re-ordered, equivalent changes made to naming or documentation, and unneeded import statements or commented out code is  preserved or removed. 
   
      One example in the study of conflicting changes that are both equally acceptable, and one is arbitrarily accepted by the resolving author is when authors of conflicting branches renamed the same variable with a slight variation:\\ \ic{SPAN\_TARGET\_ATTRIBUTE\_NAME} and \\ \ic{SPAN\_TARGET\_APP\_ID\_ATTRIBUTE\_NAME}. In these cases, either version selected by the merging algorithm might still be acceptable to the developer. \textsc{MergeBert} generated a suggestion to keep the variable name \ic{SPAN\_TARGET\_ATTRIBUTE\_NAME} whereas the user resolution originally kept the other. Participant P5 marked this resolution as acceptable and semantically equivalent, explaining that in this scenario they had `no preference as to which one is better'.

\mybox{Takeaway 1}{grey!20}{grey!7}{Evaluating the performance of \thistool{} using strict syntactic approaches estimates a lower bound of performance. Survey results show  almost 45\% of \thistool{} suggestions are acceptable merges that are semantically equivalent to the participant's original resolution. \thistool{}'s performance could be improved by considering semantic information, for example, to identify how changes related to naming or documentation should be merged.}
\noindent{}\textbf{Tangled Code Changes in Oracle} (10/122) \\
Resolutions for 10 of the conflicts contained additional ``tangled'' changes~\cite{Herzig:msr13:ImpactOfTangledCodeChanges,Kirinuki:icpc14:TangledChanges} that were unrelated to the resolution. Examples include renaming a method and adding a variable in the conflict region that is then used later outside the conflict region.
In all 10 instances, \textsc{MergeBert} generates a suggestion that does not include the additional tangled code, but is acceptable to the participant as a resolution of the conflict. 
Participants indicated that if they had access to the \textsc{MergeBert} suggestions, they would select the correct resolution and then manually add the additional code. 

\mybox{Takeaway 2}{grey!20}{grey!7}{When committing merged code, developers may introduce changes unrelated to the conflict which are inadvertently included in conflict resolution oracles. These changes can negatively impact model performance estimated with automatic metrics.}

\subsubsection{Merge Requires External Context}
 \textsc{MergeBert} did not generate an acceptable suggestion for 46\% (56/122) of examples shown to survey participants. Participants were asked to indicate whether they resolved these examples using external context that cannot be inferred from the conflicting code regions and to explain what the external context was.  
Results indicate that 16\% (20/122) of conflicts in the survey sample require external information not found in either conflicting file, in order to be correctly resolved. 
One example of external context is knowledge of linter rules enabled at a project level. Projects often require linter checks before code can be committed to the repository, as a step towards improving the quality and maintainability of the source code. One example is a merge conflict from Roslyn where the correct resolution was to remove a null check from the code. Participant P23 explained the decision to remove the check: \emph{"The previousResults parameter is non-nullable because C\# nullability checking is now enabled at the project level. The null check is unnecessary"}. In this scenario, without specific knowledge of linter checks, an automatic approach is unable to predict an accurate merge. 
 
 Another example of external context is updates to languages rules that have cascading effects on existing code. Participant P22 from the Roslyn project explained one such conflict:  "Changes were due to updates in '\ic{using}' rules for the C\# language".  Language updates in C\# version 8.0 introduced an alternative syntax for the \ic{using} statement and P22's team had made to adopt this syntax.  P22 therefore updated this code (involved in the conflict) during the merge. Other examples of external context identified through the survey include: removal of global dependencies from non-conflicting files within a project, rolling back features that shouldn't be included in a release branch, and project-level decisions to remove \ic{'final'} modifiers for variables. 

\mybox{Takeaway 3}{grey!20}{grey!7}{The local view of a conflict is sufficient to merge a majority of conflicts. Around 16\% of the conflicts  require external information to correctly resolve. One direction to improve \thistool{} is to consider external context as an additional information source for resolving conflicts.}

\subsubsection{Unacceptable Merge Suggestions}
Survey results show that \textsc{MergeBert} suggestions were incorrect for 29\% (36/122) of the conflicts. Participants indicated that none of the 36 conflicts required external context to be resolved. We manually analyze the conflicts looking to identify patterns that may explain the incorrect merges, for example, affected language construct~\cite{pan2021ProgramSynthesis} and type of conflict~\cite{shen2021automatic}, but do not identify any consistent patterns. 
In summary, existing automatic evaluation strategies estimate a lower bound of approach performance: \thistool{} suggestions are correct for 54\% of conflicts included in our sample, despite not being syntactically equivalent to the user resolution.  Further, suggestions from \thistool{} helped two participants find bugs in their own recent merge conflict resolutions!  This is in addition to those resolutions where \thistool{} does provide an exact match.  This finding suggests that automatic evaluation techniques that rely on a strict syntactic comparison between the user resolution and merge suggestion might be estimating a much lower bound of performance. This highlights a discrepancy between how approaches are typically automatically evaluated, and how developers may evaluate an approach in practice.  Researchers should consider conducting user studies to more accurately evaluate approaches when feasible. 
 Tools like \thistool{} can reduce effort and bug proneness involved in manually merging conflicts. Future studies should investigate these potential benefits. 

%% file: related_work.tex
\subsection{Related Work}

\newcommand{\etal}{et al. }
There have been multiple attempts to improve merge algorithms by restricting them to a particular programming language or a specific type of applications~\citep{mens2002state}. Typically, such attempts result in algorithms that do not scale well or have low coverage. Syntactic merge algorithms improve upon \texttt{diff3} by verifying the syntactic correctness of the merged programs. Several syntactic program merge techniques have been proposed \citep{westfechtel1991structure,Asklund1999IdentifyingCD} which are based on parse trees or ASTs and graphs. 

Apel~\etal noted that structured and unstructured merge each has strengths and weaknesses. 
They developed \textsc{FSTMerge}, a semi-structured merge, that alternates between approaches~\cite{apel2010semistructured}. 
Tavares~\etal implemented \textsc{jsFSTMerge} by adapting an off-the-shelf grammar for JavaScript to address shortcomings of \textsc{FSTMerge} and also modifying the \textsc{FSTMerge} algorithm itself~\cite{tavares2019semistructured}.
 Cavalcanti~\etal performed a large scale retrospective evaluation of semi-structure merge on over 30,000 merges and found that it can still suffer from false negatives, cases where there is actually a semantic conflict but the merge approach produces a (incorrect) resolution~\cite{cavalcanti2017evaluating}. They improve \textsc{FSTMerge} by adding ``handlers'' that check for common false negative cases (\emph{e.g.} renames, added references to modified elements) that remove these cases completely.
Le{\ss}enich noted that using AST representations works well for merge, but differencing is NP-hard due to renamings and shifted code. They propose an approach to improve performance of the \textsc{JDime} algorithm at minimal cost~\cite{lessenich2017renaming}.
Dinella~\etal take a data driven approach to the merge conflict resolution problem and introduce \textsc{DeepMerge}, a deep neural network that uses a pointer network architecture to construct the resolution from lines in the different input versions of the code~\cite{Dinella2021}.

Finally, \citet{Sousa18} explore the use of program verification to certify that a merge obeys a semantic correctness criteria, but does not help synthesize resolutions. 
On the other hand, \citet{pan-synthesis-2021} explore using program synthesis to learn repeated merge resolutions within a project. 
However, the approach is limited to a single C++ project, and only deals with restricted cases of import statements. 

\subsubsection{Empirical Studies}

Several empirical studies have investigated merge conflicts and challenges faced by developers in merge resolution. McKee \etal \cite{mckee2017software} and Nelson~\etal \cite{nelson2019life} interviewed developers and performed a follow-up survey with 162 developers to build a detailed understanding of developer perceptions regarding merge conflicts in general. They found, among other things, that complexity of the conflicting lines of code and file as a whole, the number of LOC in the conflict, and developers’ familiarity with the conflicting lines of code impact how difficult developers find a conflict to resolve.  
Brindescu \etal investigated the impact of merge conflicts and their resolutions on software quality~\cite{brindescu2020lifting,brindescu2020empirical}.   They found that 20\% of code changes resulted in a merge conflict and the code in these conflicts were twice as likely to contain bugs as other changes. Further, if the changes included semantically interacting changes, the likelihood of a defect is 26 times that of non-conflicting changes.

Costa \etal presented TIPMerge, an approach for identifying and recommending developers to participate in merge sessions when resolving conflicts~\cite{costa2016tipmerge}. They  evaluated it on 2,040 merges across 25 open source projects and found that TIPMerge can improve joint knowledge coverage by an average of 49\% in merge scenarios~\cite{de2019recommending}.

Vale et al. \citep{vale2021challenges} performed an empirical study to understand what makes merge challenging for developers.  Through a large scale automated analysis and a survey of 140 developers, they identified factors that contribute to merge conflict resolution difficulty (e.g., number of chunks in the conflict and number of developers involved in the merge scenario).
Seibt et al. \citep{seibt2021leveraging} explore and evaluate merge algorithms on a suite of ten software repositories, paying attention to the amount of resolutions produced, size of conflict, runtime cost, and correctness. Interestingly, they use the test suites of each project as an oracle to assess correctness of code after the merge.

None of the existing studies evaluate automatic merge resolution tools with software developers on their own real-world conflicts. The participants in our survey have expertise to understand when \thistool{} resolution suggestions would be acceptable on their own real-world conflicts, providing rich explanations about when external context is required, or when tangled code changes are made.

%% file: threats.tex
\section{Threats to Validity}

The choice of hyper-parameters in our model (Section ~\ref{sec:implement}) is based on prior work of others and generally accepted norms~\cite{bert}.  It's possible that exploring the hyper-parameter space could yield different results.
The sample of conflicts and projects used in the study may pose a threat to the external validity of our work. We only considered public open-source projects hosted on GitHub, therefore, results may not generalize to closed source projects or repositories hosted on other platforms. To mitigate this threat, we select a diverse set of projects varying in size and language. Similarly, survey participants evaluate their own recently-merged conflicts and the set of conflicts used in the survey to answer RQ4 may not be a representative sample, as it was dependent on participant availability. We filtered out merge conflicts from the user study that we considered to be ``trivial'' conflicts.  This was a subjective judgement, but we did aim to select substantive conflicts in the hopes that they would elicit more valuable and informative feedback from participants.
The survey interface replicates the VSCode diff3 view. Participants not familiar with this view may have a harder time navigating the conflict view and answering survey questions, to mitigate this threat, we create an instructional video for participants to watch. 
 
 

%% file: conclusion.tex
\section{Conclusion}
\balance
This paper introduces \thistool{}, a transformer-based program merge framework that leverages token-level differencing and reformulates the task of generating the resolution sequence as a classification task over a set of primitive merge patterns extracted from real-world merge commit data. \thistool{} exploits pretraining over massive amounts of code and then finetuning on specific programming languages, achieving 64--69\% precision and 63--68\% recall of merge resolution synthesis.  Lastly, \thistool{} is flexible and effective, capable of resolving more conflicts than the existing tools in multiple programming languages.

To better evaluate the resolutions generated by \thistool{} from the perspective of users, we conduct a user study with 25 developers from large OSS projects. We ask participants to evaluate whether \thistool{} resolution suggestions are acceptable on a set of 122 of their own real-world conflicts. Results suggest, in practice, \thistool{}  resolutions would likely be accepted at a higher rate than estimated by the performance metrics chosen. Using participant feedback we identify potential ways to improve \thistool{} by improving the oracle to remove tangled changes or considering external context -- project or team level information that is not present in the conflicting files.